\documentclass[fleqn,usenatbib]{mnras}

\usepackage{newtxtext,newtxmath}
\usepackage[T1]{fontenc}
\usepackage{ae,aecompl}
\usepackage{graphicx}	
\usepackage{amsmath}	
\newcommand{\edit}[1]{#1}

\title[The blue-optical transmission spectrum of WASP-74b]{A precise blue-optical transmission spectrum from the ground: Evidence for haze in the atmosphere of WASP-74b}

\author[P. Spyratos et al.]{Petros Spyratos$^{1}$\thanks{E-mail: p.spyratos@keele.ac.uk},
Nikolay K. Nikolov$^{2}$,
Savvas Constantinou$^{3}$,
John Southworth$^{1}$,
\newauthor
Nikku Madhusudhan$^{3}$,
Elyar Sedaghati$^{4}$,
David Ehrenreich$^{5}$,
and Luigi Mancini$^{6,7,8}$
\\
$^{1}$Astrophysics Group, Keele University, Staffordshire, ST5 5BG, UK\\
$^{2}$Space Telescope Science Institute, 3700 San Martin Dr, Baltimore, MD 21218, USA\\
$^{3}$Institute of Astronomy, University of Cambridge, Madingley Road, Cambridge CB3 0HA, UK\\
$^{4}$European Southern Observatory, Alonso de C\'ordova 3107, Vitacura, Regi\'on Metropolitana, Chile\\
$^{5}$Observatoire Astronomique de l'Université de Genève, Chemin Pegasi 51, Versoix, CH-1290 Versoix, Switzerland\\
$^{6}$Department of Physics, University of Rome 'Tor Vergata', Via della Ricerca Scientifica 1, I-00133, Rome, Italy\\
$^{7}$Max Planck Institute for Astronomy, Königstuhl 17, D-69117, Heidelberg, Germany\\
$^{8}$INAF, Osservatorio Astrofisico di Torino, via Osservatorio 20, I-10025, Pino Torinese, Italy\\
}

\date{Accepted XXX. Received YYY; in original form ZZZ}

\pubyear{2021}

\begin{document}
\label{firstpage}
\pagerange{\pageref{firstpage}--\pageref{lastpage}}
\maketitle

\begin{abstract}
We report transmission spectroscopy of the bloated hot Jupiter WASP-74b in the wavelength range from 4000 to 6200\,\AA. We observe two transit events with the Very Large Telescope FOcal Reducer and Spectrograph (VLT FORS2) and present a new method to measure the exoplanet transit depth as a function of wavelength. The new method removes the need for a reference star in correcting the spectroscopic light curves for the impact of atmospheric extinction. It also provides improved precision, compared to other techniques, reaching an average transit depth uncertainty of 211\,ppm for a solar-type star of V=9.8 mag and over wavelength bins of 80\,\AA. The VLT transmission spectrum is analysed both individually and in combination with published data from Hubble Space Telescope (HST) and \textit{Spitzer}. The spectrum is found to exhibit a mostly featureless slope \edit{and equilibrium chemistry retrievals with PLATON favour hazes in the upper atmosphere of the exoplanet. Free chemistry retrievals with AURA further support the presence of hazes. While additional constraints are possible depending on the choice of atmospheric model, they are not robust and may be influenced by residual systematics in the data sets.} Our results demonstrate the utility of new techniques in the analysis of optical, ground-based spectroscopic data and can be highly complementary to follow-up observations in the infrared with \edit{JWST}.
\end{abstract}


\section{Introduction} \label{sec:intro}

\edit{Transiting hot Jupiters with inflated radii are some of the most prominent targets for atmospheric characterisation. During a planetary transit, part of the starlight is transmitted through the upper layers of their extended planetary envelopes,} resulting in a wavelength-dependent variation of the planetary radius. The radius variation can provide insight into the structure and composition of the planetary atmosphere and can be explored via transmission spectroscopy. Transmission spectroscopy has gained momentum in recent years by successfully identifying spectral species, including Na, K, H$_2$O, AlO, TiO and CO$_2$, as well as the absorption and scattering signatures of clouds and hazes \edit{\citep[see e.g.][]{2002ApJ...568..377C,2008A&A...487..357S,2010Natur.468..669B,2014AJ....147..161S,2017MNRAS.468.3907K,2018Natur.557..526N,2019A&A...622A..71V,2020MNRAS.494.5449C,2021MNRAS.500.5420C,2021ApJ...913L..16C,2022arXiv220811692T}}.

Observations to date have revealed that close-in, irradiated exoplanets may experience atmospheric escape under the extremely high levels of stellar irradiation. This phenomenon is now increasingly being observed in multiple hot exoplanets across a large mass range from Jupiter-mass down to Neptune-mass planets \citep{2003Natur.422..143V,2004ApJ...604L..69V,2010A&A...514A..72L,2014ApJ...786..132K,2015Natur.522..459E,2017A&A...605L...7L,2018A&A...620A.147B,2018Natur.557...68S,2019A&A...629A..47D,2020ApJ...894...97N}. Gradual atmospheric loss of volatiles could also be visible during a planetary transit as it can cause a deeper transit depth and a longer transit duration in light curves obtained from observations made in the violet and ultraviolet wavelength regions. Furthermore, a spectroscopic analysis could potentially reveal a significant difference in the transit depth observed at blue and red wavelengths.

\subsection{The WASP-74 system}

Motivated by the potential of this technique, we performed a search for the blue-optical increase of the planet radius during the transit of exoplanet WASP-74b. The WASP-74 system \citep{2015AJ....150...18H,2019MNRAS.485.5168M,2020AJ....159..137G,2020A&A...642A..50L,2022MNRAS.512.2062B} consists of an F9 dwarf star with a mass of \edit{$\sim$}1.3\,M$_{\odot}$ and a radius of \edit{$\sim$}1.5\,R$_{\odot}$ and an inflated hot Jupiter with a mass of \edit{$\sim$}0.9\,M$_\mathrm{Jup}$ and a radius of \edit{$\sim$}1.4\,R$_{\mathrm{Jup}}$. The host star has an effective temperature of \edit{$\sim$}6000\,K and a metallicity of \edit{$\sim$}0.4 \citep{2021A&A...656A..53S}. \textit{Gaia} Data Release 3 data give slightly lower values of 5800\,K for $T_{\mathrm{eff}}$ and of 0.2 for \mbox{[M/H]} \citep{2016A&A...595A...1G,2022arXiv220800211G,2022arXiv220605989B}. The exoplanet has a high equilibrium temperature of 1900\,K, a moderate surface gravity of 10\,m\,s$^{-2}$ and completes a full orbit in 2.1\,days. A search for transit timing variations from archival, published and new data by \citet{2022MNRAS.512.2062B} revealed no trends of any kind to the time of mid-transit consistent with a stable orbit.

Previous observations of the exoplanet have been contradictory with \citet{2019MNRAS.485.5168M} reporting absorption from strong molecular absorbers in the optical and \citet{2020A&A...642A..50L} detecting a steep upward slope from the near-infrared to the blue-optical end of the transmission spectrum. \citet{2021AJ....162..271F} found no evidence of metal absorbers or extreme scattering slopes. High-resolution measurements by \citet{2022A&A...657A..36L} suggest that some metals may be present in the atmosphere and report the possible detection of atomic aluminium. \citet{2018AJ....155..156T} concluded from observations in the infrared that WASP-74b is likely to be cloudy or water-depleted. \citet{2019MNRAS.485.5168M} presented two deeper and longer-than-predicted transits obtained using the Bessell $U$-band at the Danish 1.54-m telescope. These light curves were disregarded in their analysis under the assumption that they were strongly affected by systematic effects. A large transit depth at blue-optical wavelengths could reveal the presence of an extended exosphere, which has motivated us to obtain follow-up observations at short optical wavelengths.

Given that some of the published results firmly indicated stronger absorption levels in the optical and up to the near-infrared, it should be pointed out that WASP-74 is a quiet star. \citet{2015AJ....150...18H} and \citet{2021AJ....162..271F} monitored the star photometrically and established that the star is magnetically inactive. Long term ground-based surveillance of WASP-74 found largely invariable light curves and no periodic signals down to at least 1\,mmag \citep{2015AJ....150...18H,2021AJ....162..271F}. This suggests that the radius discrepancy observed between different observations is likely not a result of variable stellar flux caused by the presence of spots and variations in their occurrence and distribution on the stellar surface. 

To observe the system, we utilised the FOcal Reducer and low resolution Spectrograph \citep[FORS2,][]{1998Msngr..94....1A} mounted on the Very Large Telescope (VLT) at the European Southern Observatory (ESO) in  Paranal, Chile. This instrument has already been widely used to characterise the atmospheres of other hot exoplanets \citep{2010Natur.468..669B,2011ApJ...743...92B,2015A&A...576L..11S,2016A&A...596A..47S,2017MNRAS.468.3123S,2017Natur.549..238S,2016A&A...587A..67L,2016ApJ...832..191N,2018Natur.557..526N,2021AJ....162...88N,2017MNRAS.467.4591G,2020MNRAS.494.5449C,2020MNRAS.497.5155W,2021MNRAS.506.2853S}.

We then analysed the data following a series of techniques and developed a new method that circumvents the requirement for a comparison star in the spectroscopic analysis. Our new method takes the best-fit model from the white light curve analysis and considers only the raw target light curves in the common-mode correction and the subsequent spectroscopic fits (see Section~\ref{sec:spectroscopic_analysis_A} for more details on common-mode correction). In some ways, this new technique presents similarities to the approach developed by \citet{2022MNRAS.510.3236P}, which we also consider in our investigation. In the essence of that method, the fit is also applied to the target light curves but differs from our own method in that the common-mode is used as a regressor to a stochastic Gaussian Process \citep{2006gpml.book.....R,2012MNRAS.419.2683G} instead of being removed by the usual linear operation.

We found that our new method significantly improves the precision of the transmission spectrum parameters compared to both the classic method (by 71\%) and the \citet{2022MNRAS.510.3236P} method (by 49\%). This precision translates to a mean transit depth uncertainty of 210\,ppm for our new method. The constructed transmission spectrum revealed a steep slope, as well as a reasonable agreement with HST results \citep{2021AJ....162..271F}. The result suggests a plausibly hazy atmosphere for WASP-74b, in accordance with findings from \citet{2020A&A...642A..50L}.

This paper is organised as follows: Section~\ref{sec:observations_and_data_reduction} outlines the observations and describes the data reduction procedure. Section~\ref{sec:analysis} presents the steps followed for the combined (white) and binned (spectroscopic) light curve analyses. Section~\ref{sec:transmission_spectrum} discusses the outcomes and the obtained transmission spectrum. Section~\ref{sec:platon} reveals results from a retrieval analysis using PLATON and section~\ref{sec:summary} summarises our conclusions on the atmosphere of WASP-74b. 

\section{Observations and Reductions}
\label{sec:observations_and_data_reduction}

We observed two complete transits of WASP-74b using FORS2 as part of ESO program 0101.C-0716 (P.I.\ Southworth). These observations were carried out on the nights of June 20th and August 19th 2018. A third transit was also observed on May 21st but was discarded from our analysis due to its low quality owing to the poor observing conditions throughout that night.

We utilised the multi-object spectroscopy mode of the instrument to collect spectra of the target and three comparison stars (2MASS J20181210$-$0107143, 2MASS J20180844$-$0102001 and 2MASS J20180810$-$0101305). We used broad slits (22$\arcsec\times$22$\arcsec$) to reduce differential slit losses caused by seeing variations. To improve the duty cycle, we used a binning of 2$\times$2, which reduced readout to $\sim30\sec$. The spectrum of the brightest comparison star (2MASS J20181210$-$0107143) was recorded on the second CCD chip whereas the two fainter comparison stars were recorded on the same chip as the target. The second CCD chip was highly corrupted by unknown systematic effects during the June observation (i.e. the data points were greatly scattered throughout most of the observation with the light curve showing significant variations from one data point to another), and so a relative flux transit light curve could not be obtained on that date for the brightest reference star. We, therefore, chose to rely on the two fainter nearby stars for our analysis. We investigated the out-of-transit data of the comparison stars (individually and combined) and found that the light curve of the second brightest reference star (2MASS J20180844$-$0102001) resulted in the least amount of scatter. Hence, we opted to use this star to correct for any atmospheric effects.

During both nights, we utilised grism GRIS600B to obtain the spectra of the target and the comparison stars. This dispersive element produces spectra covering the wavelength region from 3300 to 6200~\AA. We observed under clear weather conditions throughout most of the two nights with seeing fluctuating between 0.3$\arcsec$ and 1.7$\arcsec$, but remaining below 1$\arcsec$ in most of the recordings. A thin cirrus cloud crossed over the field during the middle of the transit for the August data set while the June data set had an integrated water vapour that was nearly three times higher compared to the August observation. Trends in water vapour and other optical state parameters can be seen in Figure~\ref{fig:optical_state_parameters}. The target and the reference stars were monitored for 4 hours and 30\,$\min$ on the first night covering airmass from 1.09 to 1.50, and for 5 hours and 15\,$\min$ on the second night covering airmass from 1.09 to 1.47. Data collection was suspended for $\sim$7\,$\min$ after the transit in June due to technical issues. Since these data were out-of-transit and sufficient data had been collected for the time before, during and after the transit, we decided to remove all data after the observed gap in the recording. This resulted in 333 and 378 exposures, respectively, with integration times spanning from 16\,s to 35\,s.

\begin{figure*}
\centering
\includegraphics[width=0.75\textwidth,height=0.75\textheight,keepaspectratio]{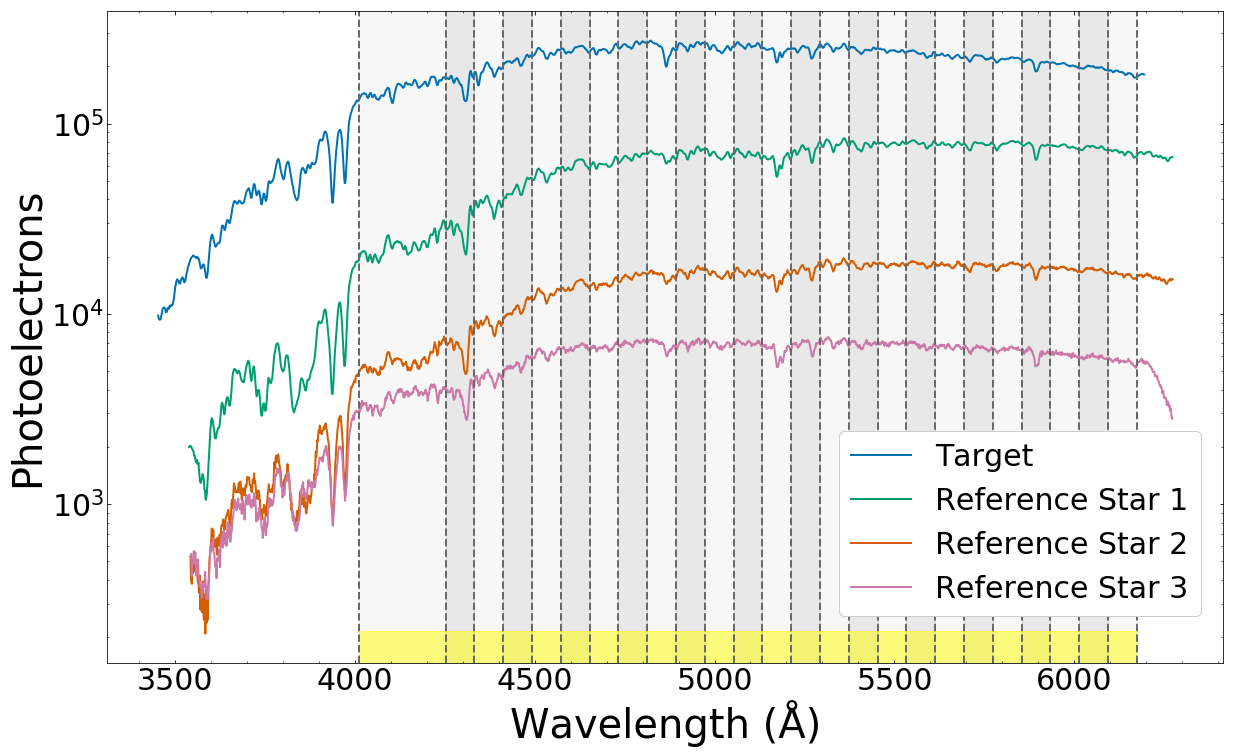}
\caption{\edit{Example spectra of WASP-74 and three reference stars. The blue line indicates the target whereas green, brown, and magenta represent three reference stars. The shaded grey bands represent the spectroscopic channels and the yellow region shows the wavelength range used for the combined light analysis.}}
\label{fig:spectra}
\end{figure*}

The data were reduced using the methods detailed in \citet{2021MNRAS.506.2853S} and \citet{2016ApJ...832..191N,2018Natur.557..526N}. When performing spectral extraction, we used an aperture radius of 15 pixels for both transit observations as it minimised the observed scatter in the out-of-transit data. \edit{Furthermore, we specified background regions from 25 to 30 pixels away on both sides of the middle of the spectral trace for each column. The pixels from both  background regions were used to compute a median value, which was then subtracted from the pixels included in the extraction aperture for each column of the two-dimensional spectra. Example one-dimensional spectra from the target and the three comparison stars are shown in Fig.~\ref{fig:spectra}.}

We produced spectrophotometric light curves covering the wavelength region between 4013 and 6173\,\AA. In particular, we created 25 narrow spectroscopic bins of 80\,\AA\ in width to explore the planet-to-star radius ratio as a function of wavelength, and one large, white channel to get the initial system parameters. We adopted a bin with a wider width of 240\,\AA\ for the wavelengths between 4013 and 4253\,\AA\ to compensate for the lower signal-to-noise ratio at these wavelengths. We did this for the target and the comparison stars of each data set. We then obtained the \edit{white} light curves from the combined light of the entire wavelength range investigated by dividing the summed flux of the target by the summed flux of the second brightest reference star. These relative fluxes were used in the white light curve analysis (Section~\ref{sec:white_analysis}). We also constructed differential light curves for each bin, which we used in our initial (classic) approach (Section~\ref{sec:spectroscopic_analysis_A}).

\section{Data Analysis}
\label{sec:analysis}

\begin{figure*}
\centering
\includegraphics[width=\textwidth,height=\textheight,keepaspectratio]{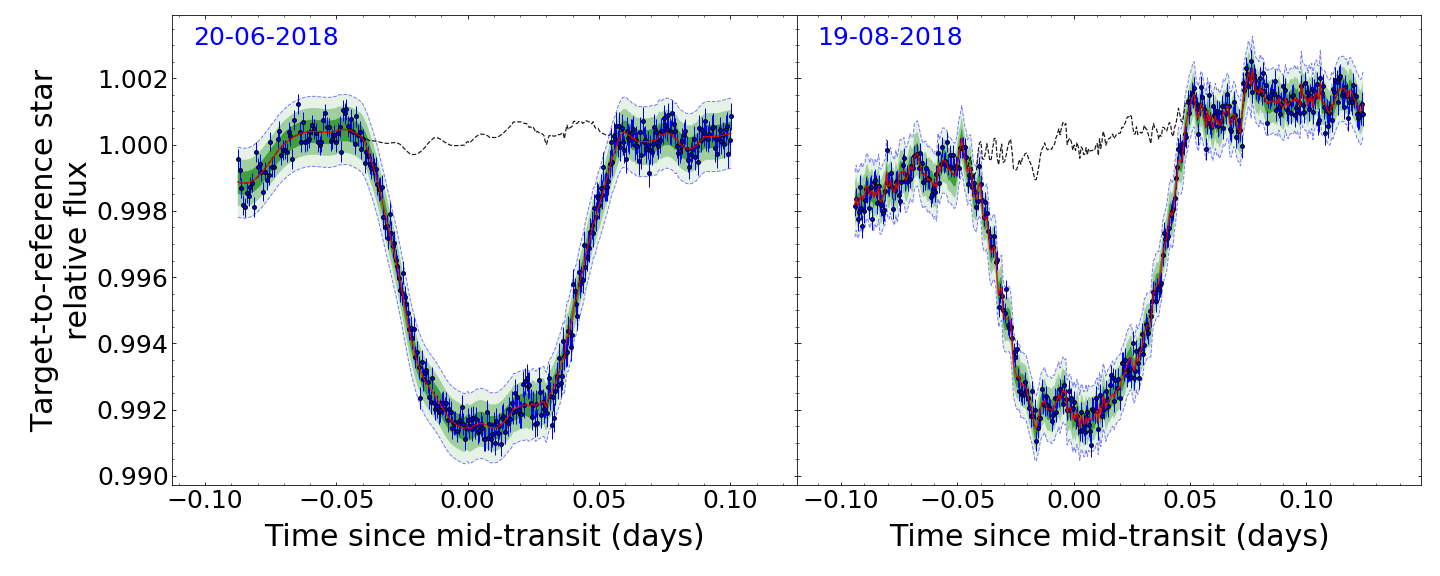}
\caption{White transit light curves of WASP-74b. The two data sets (blue dots) are plotted with their 1$\sigma$ \edit{spectrophotometric errors increased by the multiplicative factor $\beta$} (vertical bars). The black dashed line represents the noise model and the red line indicates the GP fit. The transparent green areas illustrate the 1, 2 and 3$\sigma$ scatter of the residuals from darker to lighter shades, respectively.}
\label{fig:both_white}
\end{figure*}

\begin{table}
\centering
\caption{WASP-74b system and noise parameters.}
\label{tab:wasp74_white}
\begin{tabular}{lcc}
\hline
\hline
Parameter & Value & Prior\\
\hline
Period (d) & 2.1377445 (fixed)\\
Eccentricity & 0 (fixed)\\
$a/R_*$ & 4.82 (fixed)\\[2pt]
$i$ (degrees) & 79.86 (fixed)\\[2pt]
$u_2$ & 0.29 (fixed)\\[2pt]
 & \\
20-06-2018 & \\
$\ln\alpha$ & $-15.41^{+0.57}_{-0.46}$ & $\mathcal{U}$(-20,15)\\[2pt]
$\ln\tau_{\mathrm{x}}$ & $-3.54^{+0.41}_{-0.49}$ & $\mathcal{U}$(-15,15)\\[2pt]
$c_0$ & $1.00008^{+0.00016}_{-0.00017}$ & $\mathcal{U}$(0.9,1.1)\\[2pt]
$c_1$ & $0.000087^{+0.000084}_{-0.000088}$ & $\mathcal{U}$(-0.1,0.1)\\[2pt]
$t_0$ (BJD$_{\rm TDB}$) & 2458289.77937 $\pm$ 0.00038 & $\mathcal{U}$(-0.01,0.01)$^{*}$\\[2pt]
$R_\mathrm{p}/R_*$ & $0.09883^{+0.00072}_{-0.00074}$ & $\mathcal{U}$(0.03,0.15)\\[2pt]
$u_1$ & $0.420^{+0.040}_{-0.044}$ & $\mathcal{U}$(0,1)\\[2pt]
$\mathrm{\beta}$ & $1.546^{+0.077}_{-0.091}$ & $\mathcal{U}$(0,10)\\[2pt]
 & \\
19-08-2018 & \\
$\ln\alpha$ & -15.81 $\pm$ 0.21 & $\mathcal{U}$(-20,15)\\[2pt]
$\ln\tau_{\mathrm{x}}$ & $-6.21^{+0.50}_{-0.31}$ & $\mathcal{U}$(-15,15)\\[2pt]
$\ln\tau_{\mathrm{y}}$ & $-0.49^{+0.55}_{-0.43}$ & $\mathcal{U}$(-15,15)\\[2pt]
$c_0$ & $1.000265^{+0.000070}_{-0.000074}$ & $\mathcal{U}$(0.9,1.1)\\[2pt]
$c_1$ & 0.000897 $\pm$ 0.000046 & $\mathcal{U}$(-0.1,0.1)\\[2pt]
$t_0$ (BJD$_{\rm TDB}$) & 2458349.63287 $\pm$ 0.00023 & $\mathcal{U}$(-0.01,0.01)$^{*}$\\[2pt]
$R_\mathrm{p}/R_*$ & 0.0951 $\pm$ 0.0013 & $\mathcal{U}$(0.03,0.15)\\[2pt]
$u_1$ & $0.411^{+0.043}_{-0.046}$ & $\mathcal{U}$(0,1)\\[2pt]
$\mathrm{\beta}$ & $1.410^{+0.112}_{-0.096}$ & $\mathcal{U}$(0,10)\\[2pt]
\hline
\multicolumn{3}{p{\linewidth}}{$^{*}$The prior of the time of mid-transit is set around the expected value from the ephemeris reported in \citet{2019MNRAS.485.5168M}.}
\end{tabular}
\end{table}

\subsection{White light curves}
\label{sec:white_analysis}

The white light curves show that the two data sets do not exhibit significant flux variations, indicating high atmospheric transparency. Nevertheless, the systematics were still substantial and so we chose to handle them stochastically by using a Gaussian Process (GP) framework \citep{2012MNRAS.419.2683G} to extract the transit parameters. 

A transit light curve can, thus, be expressed by a multivariate normal distribution that consists of two parts. One is the stochastic noise component that is represented by a Mat\'{e}rn 3/2 kernel, which we use throughout this work, and the other is the deterministic mean function that defines the transit. In this study, we use the Python GP toolbox \texttt{george} \edit{\citep{2015ITPAM..38..252A,2015ascl.soft11015F}} to disentangle the noise from the transit light curves. \edit{The covariance function $K_{n m}$, assumed in our study, is given by:}
\begin{eqnarray}
K_{n m} & = & \xi^{2}\left(1+\sqrt{3\sum_{\nu=1}^{N}\left(\frac{\Delta\hat{w}_{\nu}}{\tau_{w_\nu}}\right)^{2}}\right) \exp \left(-\sqrt{3\sum_{\nu=1}^{N}\left(\frac{\Delta\hat{w}_{\nu}}{\tau_{w_\nu}}\right)^{2}}\right) \nonumber \\
 &  & + \,\,\delta_{n m} \left(\sigma_{p} \beta\right)^{2},
\end{eqnarray}
where $\xi$ is the characteristic amplitude or height scale, $\tau_{w_\nu}$ are the characteristic length scale parameters for the noise variables $\hat{w}_{\nu}$ considered, $\delta_{n m}$ is the Kronecker delta, and $\sigma_{p}$ are the spectrophotometric shot noise uncertainties multiplied by a constant factor $\beta$. The circumflex symbols indicate the external systematic variables that are set to the same scale through subtraction of the mean and division by the standard deviation.

In our analysis, the mean function is a product of a model transit light curve and a linear function of time. To compute the model transits, we utilised the open source python package \texttt{batman} \citep{2015PASP..127.1161K} assuming a quadratic limb darkening law \citep{1950HarCi.454....1K} and the analytic formulae provided by \citet{2002ApJ...580L.171M}. Throughout our work, time refers to the central exposures times, which were converted from Modified Julian Dates to Barycentric Julian Dates (BJD) using the Python software \texttt{barycorrpy} \citep{2018RNAAS...2a...4K}.

We examined various external noise variables ($\hat{w}_{\nu}$) to determine the kernel function that best describes each data set. We explored several noise factors, including spectral shifts, changes in the rotator angle, FWHM, and time. These GP kernel inputs could contribute to the shape of the light curves either separately or in combination. We found that a kernel function of the rotator angular velocity ($x$) for June, and a kernel function of the rotator angular velocity and the displacement in the cross-dispersion axis ($y$) for August led to low residual scatter \edit{(353 and 328\,ppm, respectively)} and were therefore selected for the white light curves. \edit{We examined results for the transmission spectra obtained assuming different noise configurations and found good agreement between these results (i.e. the produced transmission spectra were very similar in shape and transit depth uncertainties). Thus, we determined that} the choice of regressors in the white analysis had a negligible effect on the final result \edit{that was obtained based on the methodology presented in Section~\ref{sec:spectroscopic_analysis_D}}.

To compute the best-fit kernel and transit parameters and obtain their uncertainties, we followed the same Markov Chain Monte Carlo (MCMC) sampling procedure outlined in \citet{2021MNRAS.506.2853S}. We used the default ensemble sampler included in the python package \texttt{emcee} \citep{2013PASP..125..306F}, which explores the parameter space through a set of walkers that gradually move towards the maximised likelihood. To get the optimised parameters, we performed one fit that consisted of two iterations. The iterations involved two burn-in phases with 150 walkers and 500 steps each, and one production phase with the same amount of walkers and 2000 steps. Before the first phase, the walkers for the transit parameters were initialised in a limited space around the reported or expected values from \citet{2019MNRAS.485.5168M}. These walkers were re-initialised before the second burn-in phase to a narrow zone around the location of the walker with the best probability to expedite convergence towards the maximum likelihood.

We allowed three transit parameters ($t_0$, $R_\mathrm{p}/R_*$, $u_1$), a multiplicative factor $\beta$, two or three kernel parameters ($\xi$, $\tau_{w_x}$, $\tau_{w_y}$), and a linear trend, parameterized with $c_0$ and $c_1$, to run as free parameters during the white transit light curve MCMC fit. The semi-major axis to stellar radius ratio ($a/R_*$) and inclination ($i$), were fixed to the measurements reported in \citet{2019MNRAS.485.5168M}, to enable a direct comparison of the VLT transmission spectrum with the spectrum from HST reported by \citet{2021AJ....162..271F}, who made use of the same system parameters. We also fixed the orbital period to the value reported by \citet{2019MNRAS.485.5168M} and assumed a circular orbit. The time of mid-transit ($t_0$) was initially set to the predicted value from the ephemeris given in \citet{2019MNRAS.485.5168M} and the planet-to-star radius ratio ($R_\mathrm{p}/R_*$) was also placed to the values reported there. Furthermore, the quadratic limb darkening coefficients were produced from synthetic spectra of 3D model stellar atmospheres. For their estimation, we employed the Stagger-grid \citep{2015A&A...573A..90M} and used the closest match to the temperature, surface gravity and metallicity presented in \citet{2015AJ....150...18H}. We let the linear limb darkening coefficient ($u_1$) vary freely, and kept the quadratic one ($u_2$) fixed. We used log-uniform priors for the kernel parameters and uniform priors for all other parameters. Another iteration was performed after the removal of any outliers located further than three times the standard deviation of the light curve residuals. The median GP fit parameters from the marginalised posterior of the second iteration are shown in Table~\ref{tab:wasp74_white} and the median GP model along with the systematics model and the residual errors are shown in Fig.~\ref{fig:both_white}.

\subsection{Spectroscopic light curves}
\label{sec:spectroscopic_analysis}
We analysed the spectroscopic light curves following the methodologies listed below and in Table~\ref{tab:methods}, aiming at the most efficient removal of systematic effects and the highest precision in the transit depths. We considered four approaches:
\begin{itemize}
\item the classic method, where relative (target-to-reference star) spectroscopic light curves are produced, then common-mode corrected and modelled with a GP kernel of time, 
\item the \citet{2022MNRAS.510.3236P} method, where the raw spectroscopic fluxes of the target are fitted with a GP of the common-mode and time, 
\item a modified version of the \citet{2022MNRAS.510.3236P} method, where the mean function is described by the product between the transit model and an exponential of airmass, and 
\item our new method, where the common mode is computed by dividing the raw white light curve of the target by the best-fit transit model, and then the raw spectroscopic light curves of the target are divided by the common-mode for the relevant observation and fitted with a GP that includes a regressor of time and a mean function of the transit model multiplied by the exponential of airmass. 
\end{itemize}

\begin{table*}
\centering
\caption{A summary of the approaches considered for the analysis of the spectroscopic light curves. The initials T and E stand for the transit model and the airmass exponential, respectively.}
\label{tab:methods}
\begin{tabular}{llcccc}
\hline
\hline
Method & Light curve & CM$^a$ correction & GP regressors & Mean function & $\bar{\sigma_{\rm \delta}}$ (ppm)$^b$\\
\hline
Our new method & Target & Yes & time & T $\times$ E & 211\\
Classic method & Target/Reference & Yes & time & T & 719\\
\citet{2022MNRAS.510.3236P} & Target & No & time, CM & T & 533\\
Modified \citet{2022MNRAS.510.3236P} & Target & No & time, CM & T $\times$ E & 412\\
\hline
\multicolumn{6}{p{16cm}}{$^a$CM: common-mode, $^b\bar{\sigma_{\rm \delta}}$: mean transit depth $\delta=(R_{\rm p}/R_*)^2$ uncertainty in parts-per-million (ppm)}
\end{tabular}
\end{table*}

\subsubsection{The classic approach: target-to-reference star relative flux}
\label{sec:spectroscopic_analysis_A}

\begin{table}
\centering
\caption{Transmission spectrum and limb darkening coefficients from the combined data set of WASP-74b.}
\label{tab:wasp74_spectroscopic}
\begin{tabular}{lccc}
\hline
\hline
Wavelength Range (\AA) & $R_\mathrm{p}/R_*$ & $u_1$ & $u_2$\\
\hline
$4013-4253$ & $0.09650^{+0.00136}_{-0.00138}$ & $0.641^{+0.021}_{-0.021}$ & 0.195\\[2pt]
$4253-4333$ & $0.09705^{+0.00164}_{-0.00144}$ & $0.627^{+0.020}_{-0.021}$ & 0.156\\[2pt]
$4333-4413$ & $0.09610^{+0.00127}_{-0.00133}$ & $0.487^{+0.040}_{-0.044}$ & 0.240\\[2pt]
$4413-4493$ & $0.09838^{+0.00162}_{-0.00138}$ & $0.549^{+0.027}_{-0.027}$ & 0.259\\[2pt]
$4493-4573$ & $0.09685^{+0.00089}_{-0.00079}$ & $0.511^{+0.020}_{-0.022}$ & 0.262\\[2pt]
$4573-4653$ & $0.09737^{+0.00091}_{-0.00093}$ & $0.555^{+0.019}_{-0.020}$ & 0.250\\[2pt]
$4653-4733$ & $0.09646^{+0.00086}_{-0.00070}$ & $0.488^{+0.015}_{-0.014}$ & 0.259\\[2pt]
$4733-4813$ & $0.09513^{+0.00086}_{-0.00099}$ & $0.469^{+0.022}_{-0.024}$ & 0.273\\[2pt]
$4813-4893$ & $0.09742^{+0.00273}_{-0.00267}$ & $0.364^{+0.062}_{-0.066}$ & 0.356\\[2pt]
$4893-4973$ & $0.09469^{+0.00076}_{-0.00073}$ & $0.397^{+0.033}_{-0.034}$ & 0.312\\[2pt]
$4973-5053$ & $0.09390^{+0.00059}_{-0.00062}$ & $0.417^{+0.017}_{-0.016}$ & 0.274\\[2pt]
$5053-5133$ & $0.09453^{+0.00080}_{-0.00080}$ & $0.394^{+0.016}_{-0.018}$ & 0.285\\[2pt]
$5133-5213$ & $0.09457^{+0.00083}_{-0.00080}$ & $0.408^{+0.016}_{-0.015}$ & 0.287\\[2pt]
$5213-5293$ & $0.09482^{+0.00089}_{-0.00097}$ & $0.349^{+0.020}_{-0.021}$ & 0.316\\[2pt]
$5293-5373$ & $0.09555^{+0.00129}_{-0.00126}$ & $0.401^{+0.047}_{-0.049}$ & 0.312\\[2pt]
$5373-5453$ & $0.09168^{+0.00125}_{-0.00135}$ & $0.258^{+0.041}_{-0.048}$ & 0.356\\[2pt]
$5453-5533$ & $0.09495^{+0.00060}_{-0.00051}$ & $0.303^{+0.017}_{-0.017}$ & 0.311\\[2pt]
$5533-5613$ & $0.09233^{+0.00110}_{-0.00118}$ & $0.293^{+0.026}_{-0.028}$ & 0.321\\[2pt]
$5613-5693$ & $0.09462^{+0.00162}_{-0.00186}$ & $0.357^{+0.041}_{-0.041}$ & 0.319\\[2pt]
$5693-5773$ & $0.09371^{+0.00090}_{-0.00101}$ & $0.275^{+0.028}_{-0.029}$ & 0.339\\[2pt]
$5773-5853$ & $0.09362^{+0.00089}_{-0.00094}$ & $0.288^{+0.026}_{-0.026}$ & 0.311\\[2pt]
$5853-5933$ & $0.09608^{+0.00090}_{-0.00079}$ & $0.256^{+0.040}_{-0.034}$ & 0.340\\[2pt]
$5933-6013$ & $0.09358^{+0.00117}_{-0.00119}$ & $0.270^{+0.033}_{-0.033}$ & 0.344\\[2pt]
$6013-6093$ & $0.09363^{+0.00073}_{-0.00076}$ & $0.214^{+0.025}_{-0.025}$ & 0.340\\[2pt]
$6093-6173$ & $0.09384^{+0.00133}_{-0.00107}$ & $0.248^{+0.026}_{-0.027}$ & 0.320\\[2pt]
\hline
\end{tabular}
\end{table}

Under this classic approach, we applied a common mode correction to the relative, target-to-comparison star, spectroscopic fluxes, following the steps of previous FORS2 analyses \citep[e.g.][]{2016ApJ...832..191N,2018Natur.557..526N,2021A&A...656A..53S}. The common-mode was simply determined by dividing the white relative flux by the median transit GP fit. The relative spectroscopic light curves were then divided by this common trend to remove wavelength-invariant systematic effects.

Following the common-mode correction, the spectroscopic light curves were analysed under a GP framework similar to the one described in the white analysis. During this process, any additional interference was assumed to be a function of time as any significant systematics due to instrumental effects were accounted for in the common-mode correction step. The transit parameters were then computed by obeying the same fitting procedure outlined in Section~\ref{sec:white_analysis} for the white light curves.

Parameters that were initially allowed to vary were the transit parameters $R_\mathrm{p}/R_*$ and $u_1$, the multiplicative factor $\beta$, and the noise parameters $\ln\alpha$ and $\ln\tau_{\mathrm{t}}$. Time was now considered as part of the GP kernel and, thus, a linear function of time was not included. These parameters were fit separately for each spectroscopic light curve. The remaining parameters were fixed to the values acquired from the white light curve analysis and were, therefore, the same for all spectroscopic channels. \edit{A total of two outliers were identified and removed, and the fit was repeated in a second iteration.}

\begin{figure*}
\centering
\includegraphics[width=\textwidth,height=0.9\textheight,keepaspectratio]{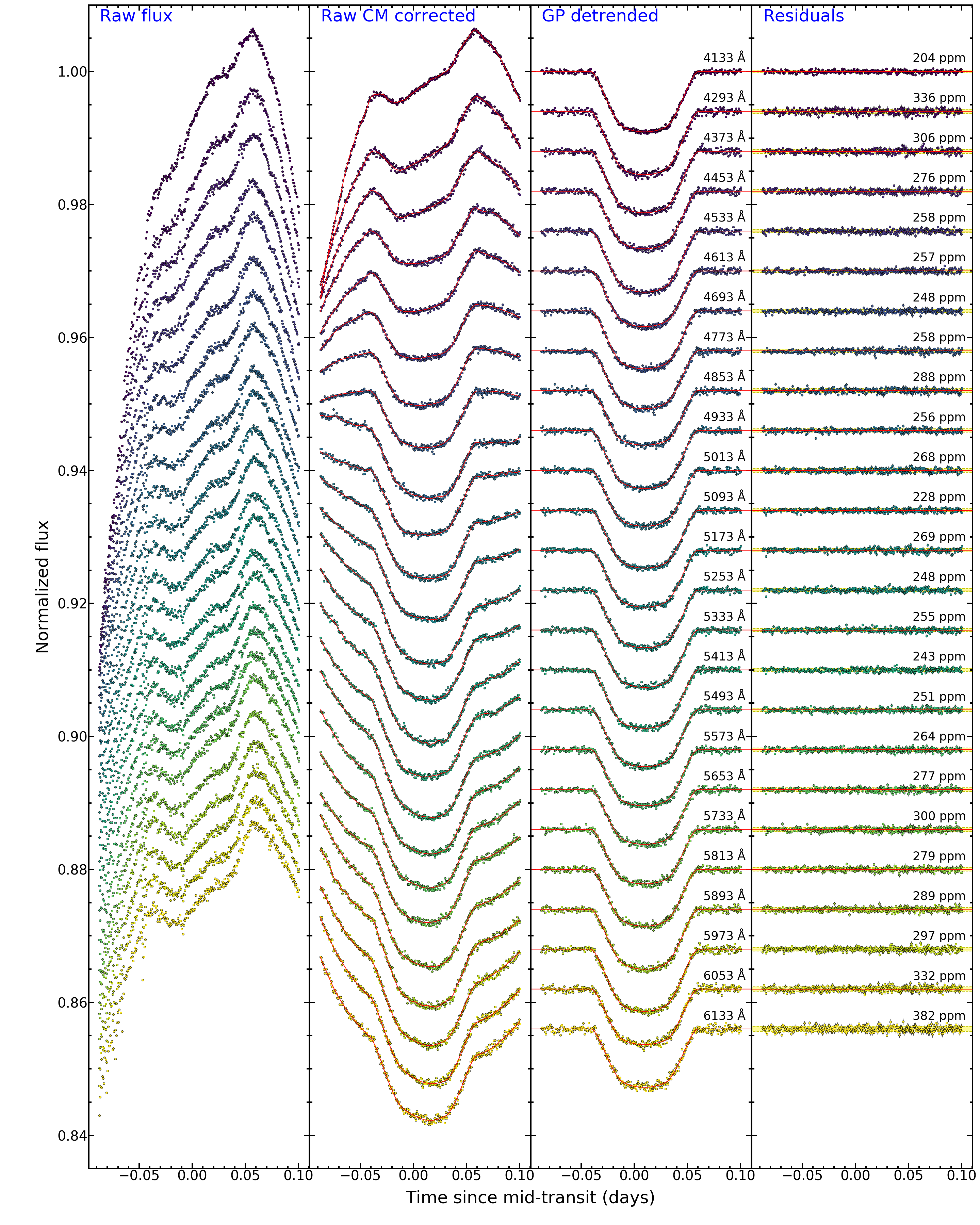}
\caption{Spectroscopic light curves for the WASP-74b data set from June 20th 2018 from our new method. Data are indicated by dots and the relevant 1$\sigma$ spectrophotometric uncertainties rescaled by a factor $\beta$ are displayed by vertical error bars. Shorter wavelengths are plotted in dark blue whereas longer wavelengths are shown in yellow. For clarity purposes, the light curves are offset from their original out-of-transit flux value of $\sim$1. First panel: Raw target fluxes. Second panel: Raw fluxes after being divided by the common mode from the raw white light curves. The GP model is indicated with a red continuous line. Third panel: Fluxes after decorrelation, and median transit model. Fourth panel: Best-fit residuals and their 1$\sigma$ residual region (yellow shaded area).}
\label{fig:one_spectroscopic}
\end{figure*}

\begin{figure*}
\centering
\includegraphics[width=\textwidth,height=0.9\textheight,keepaspectratio]{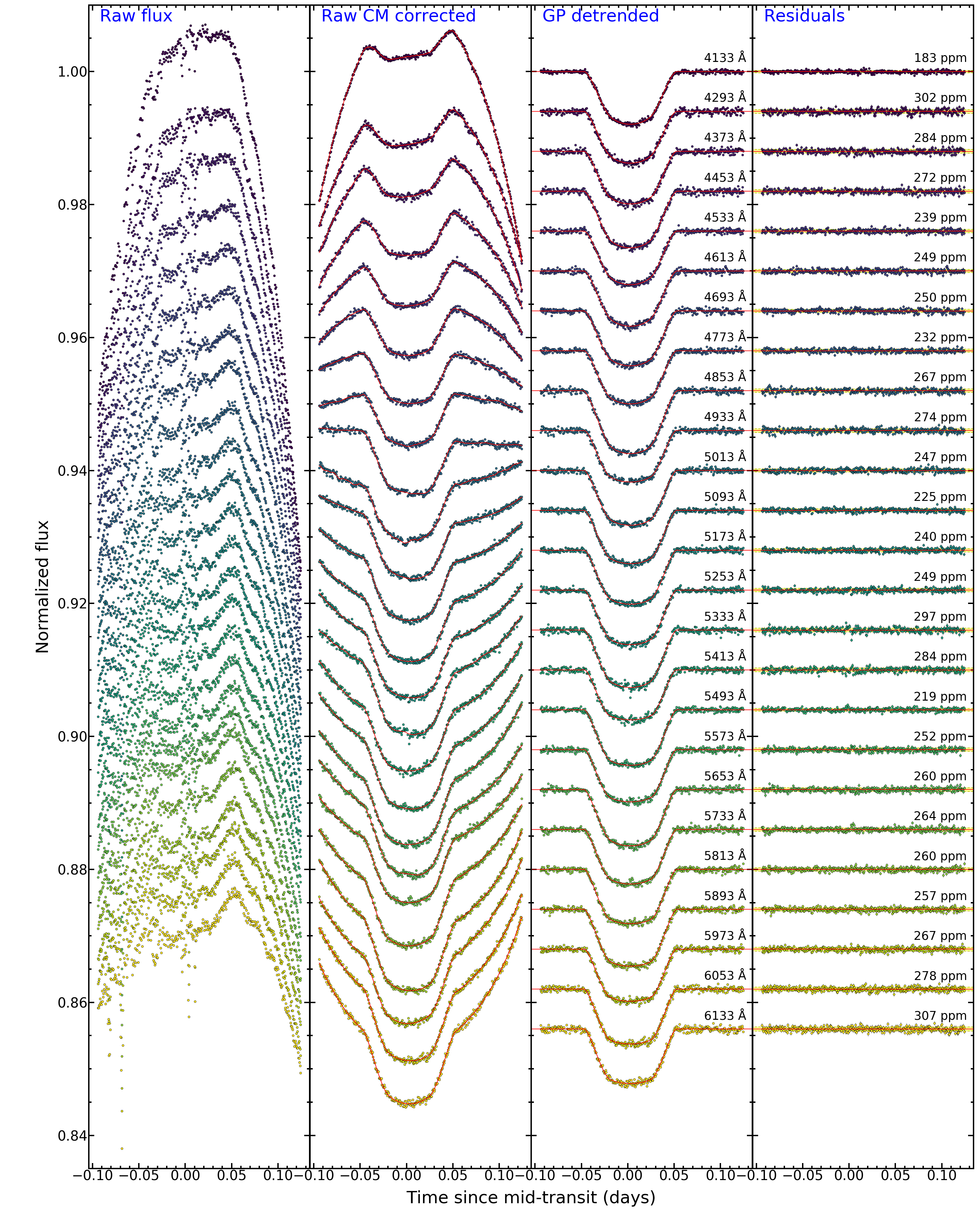}
\caption{Same as Figure~\ref{fig:one_spectroscopic}, but for the WASP-74b data set from August 19th 2018.}
\label{fig:two_spectroscopic}
\end{figure*}

\begin{figure}
\centering
\includegraphics[width=0.45\textwidth,height=0.45\textheight,keepaspectratio]{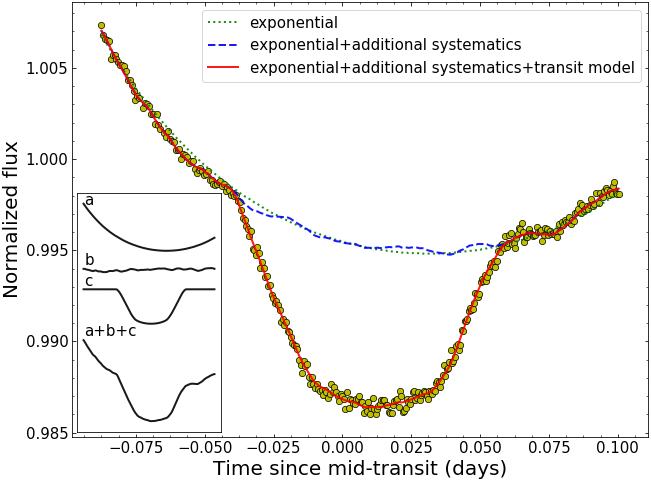}
\caption{\edit{An example of a spectroscopic light curve fitted with a GP based on our new method. On the lower left, a smaller subplot depicts each individual contribution separately (from top to bottom) and then combines them for the final result. Here, `a' represents the exponential of airmass, `b' represents the additional systematics, modelled by the GP of time, and `c' represents the transit model.}}
\label{fig:example_light_curve}
\end{figure}

\subsubsection{The \citet{2022MNRAS.510.3236P} approach: target raw flux and assuming common-mode as a regressor}
\label{sec:spectroscopic_analysis_B}

A major limitation of the classic method, in the case that the reference stars are significantly fainter than the target, is the propagation of scatter from the reference to the target light curves. This additional scatter is visible in the relative, common-mode corrected, spectrophotometric light curves, and can introduce more systematic effects. This leads to an increase in the uncertainties of the parameters obtained from fits of such light curves. Specifically, a comparison between the uncertainties of the fitted relative radii of the VLT and HST data sets, shows excessive uncertainty, over a factor of five for the VLT spectrum, when using this classic approach. Part of this lower precision may also be attributed to atmospheric effects due to the non-photometric conditions at the time of the observations and to differential effects caused by colour differences between the two stars.

To solve this issue, we ignored the flux from the comparison star completely in the spectroscopic fits and only used the raw light curves from the target. We also provided the common-mode from the raw white light curves as a regressor to the GP, following the method detailed in \citet{2022MNRAS.510.3236P}. According to this approach, the common-mode does not need to be used in the usual way to linearly correct the spectroscopic light curves, but can instead be included in the GP function as a regressor. The new kernel function was therefore modified to include both time and common mode as noise decorrelation factors. 

We then implemented a GP fit on the raw spectroscopic light curves from the target, while adhering to the same recipe outlined in Section~\ref{sec:spectroscopic_analysis_A}. We allowed the same transit and noise parameters to vary freely in the fit, plus one additional noise parameter that described the common-mode. This adaptation resulted in fits that did not require the need for a comparison star and, remarkably, increased \edit{the} precision in the obtained transmission spectrum \edit{by 26\%} (see Table~\ref{tab:methods}). \edit{We note that a similar improvement in precision was also achieved by \citet{2022MNRAS.510.3236P} for ground-based data of HAT-P-26b.}

\subsubsection{Our modification to the \citet{2022MNRAS.510.3236P} approach: taking extinction into account}
\label{sec:spectroscopic_analysis_C}

As starlight enters the Earth's atmosphere, it interacts with atmospheric constituents and so part of it is effectively blocked before reaching the Earth's surface. This phenomenon is known as atmospheric extinction and is described by the Beer-Lambert Law. The raw light curves therefore assume a characteristic form that is defined by an exponential curvature. Other physical factors may also affect the shape of these light curves, but the contribution from atmospheric extinction tends to be dominant. The index of the exponential is a function of the optical depth, which, in its simplest form, is expressed as the product between airmass and a coefficient that determines extinction. In reality, this coefficient is the sum of various coefficients that describe absorption and scattering events that reduce the incident light as it travels through the atmosphere. The nature of the extinction coefficient can therefore be quite complex, but an exact derivation of each absorption and scattering coefficient is not required for our purposes as extinction can be evaluated by one free parameter in the fit. The extinction coefficient is also wavelength-dependent, resulting in variation in the curvatures for each spectroscopic light curve.

Parametric functions of airmass are widely used in high-resolution transmission spectroscopy to perform telluric corrections. For example, one of the first studies that used such a function for this purpose was by \citet{2010A&A...523A..57V}. Their study was able to determine Earth's optical transmission spectrum based on observations of a lunar eclipse. Other studies soon followed by applying telluric corrections in this manner to high-resolution atmospheric observations of exoplanets. For instance, \citet{2013A&A...557A..56A} used an exponential function of airmass to correct for extinction effects and detect calcium and possibly other elements in the atmosphere of HD 209458b, whereas \citet{2015A&A...577A..62W} did the same to detect sodium in the atmosphere of HD 189733b.

On this basis, we deduced that the raw, target light curves should be described by an exponential function of airmass. We, therefore, adjusted our mean function to include this additional parametric function. This means that the mean function $\mu$ was now described by: \begin{equation}
\mu\left(t, z ; \mathrm{a}_{0}, \mathrm{a}_{1}, \theta\right)=\left[\mathrm{a}_0\mathrm{e}^{-\mathrm{a}_1 z}\right] \mathrm{T}(t ; \theta),
\end{equation} where $t$ is time in BJD, $z$ is the airmass, and a$_0$ and a$_1$ are coefficients that define the exponential trend. $\mathrm{T}$ represents the transit model expressed by the transit parameters $\theta$. a$_0$ is expected to be around unity since the light curves are normalised and a$_1$, the extinction coefficient, is anticipated to settle at small values.

We also adapted the fitting process and inserted an additional step before the GP fit. The initial step included performing a Levenberg-Marquardt fit to the out-of-transit data in order to obtain the initial guesses for the exponential parameters a$_0$ and a$_1$. We found that getting initial values for the airmass exponential trend aided the GP fit towards the best solution for the system. This non-linear least squares fit was carried out by making use of the python package \texttt{lmfit} \citep{2016ascl.soft06014N}. \edit{During this process, five to six deviating data points were removed per spectroscopic light curve for the August data set due to cloud effects.} 

We then executed a slightly more complex GP fit on the entire transit light curves. This fit comprised a GP kernel of time and the two-component mean function formulated earlier that consists of the transit model and the exponential of airmass. The new configuration enabled tighter constraints, especially towards the edges of the spectrum where the scatter tended to be higher. This resulted in greatly reduced error bars for $R_\mathrm{p}/R_*$, with an average decrease of over 40\% compared to the initial analysis. We, therefore, find that the inclusion of an exponential in the mean function improves $R_\mathrm{p}/R_*$ precision in the method of \citet{2022MNRAS.510.3236P}.

\subsubsection{Our new approach}
\label{sec:spectroscopic_analysis_D}

With the application of the \citet{2022MNRAS.510.3236P} prescription and the inclusion of a parametric function of airmass, we achieved a substantial reduction of the uncertainties by nearly half. Despite this promising result, a deviation in precision still remained compared to the results from the HST \citep{2021AJ....162..271F}. The greater than double size of the error bars implied that the potential to probe the planetary atmosphere was affected considerably and this rendered atmospheric retrievals less informative.

Another way to tackle this problem was to perform the usual common-mode correction technique but only to the light curves of the target. To achieve this, we first obtained the common-mode by dividing the raw, white transit light curve of the target by the median transit model from the relative white flux analysis. We then divided the raw, spectroscopic light curves of the target by this common-mode to get the common-mode corrected fluxes. We applied such a correction to each of the two data sets and then proceeded to the fits.

By following a similar procedure to the one described in Section~\ref{sec:spectroscopic_analysis_C}, we first modelled the out-of-transit data using an exponential function of airmass and performed a Levenberg-Marquardt minimisation procedure. \edit{During this procedure, three outliers were identified and discarded.} From there, we obtained initial guesses for the two systematic parameters of the parametric function. The initial assumptions for the transit parameters were acquired from the estimates in the white light analysis and the theoretical values for the limb darkening coefficients of each spectroscopic light curve were taken again from the Stagger-grid spectra. We then performed GP fits of time to the full time-series of the common-mode corrected target light curves, taking into account that the mean function was once more described by a transit model multiplied by an exponential of airmass. 

The transit parameters $R_\mathrm{p}/R_*$ and $u_1$ and all noise parameters were allowed to vary freely throughout the MCMC sampling process, whereas $t_0$, $a/R_*$, $i$ and $u_2$ were maintained at fixed values as before. This method reduced the size of the $R_\mathrm{p}/R_*$ error bars by a further $\sim$ 50\%. The various fitting stages of this analysis, as well as the best-fit residuals are shown in Figs.~\ref{fig:one_spectroscopic} and~\ref{fig:two_spectroscopic}, and the transmission spectrum parameters $R_\mathrm{p}/R_*$, $u_1$ and $u_2$ are presented in Table~\ref{tab:wasp74_spectroscopic}. \edit{In addition, Fig.~\ref{fig:example_light_curve} shows an example GP fit of a spectroscopic light curve from our new method and indicates the contribution of each separate part of the model to the final fit.}

As illustrated in Figs.~\ref{fig:one_spectroscopic} and~\ref{fig:two_spectroscopic}, the shape of the spectrophotometric light curves after common-mode correction takes a convex-flat-concave form (from top to bottom). This is because, in reality, the common-mode corrected light curves are a ratio of exponentials with different extinction indices. More specifically, it is a ratio between the exponential from the white light curve and the exponential from each spectroscopic light curve. Scattering is higher in the blue and therefore the blue light curves largely preserve the exponential form of their extinction energy distribution when divided by the common-mode. This is not the case in the red, where scattering is less and the exponential extinction from the common-mode dominates over the red light curves inverting their shape. The disparity in the extinction indexes is nearly zeroed in the mid-wavelength transition zone where the light curves obtain a nearly flat shape. In our analysis, we assume this ratio as one coefficient to simplify and accelerate calculations.

\begin{figure*}
\centering
\includegraphics[width=\textwidth,height=\textheight,keepaspectratio]{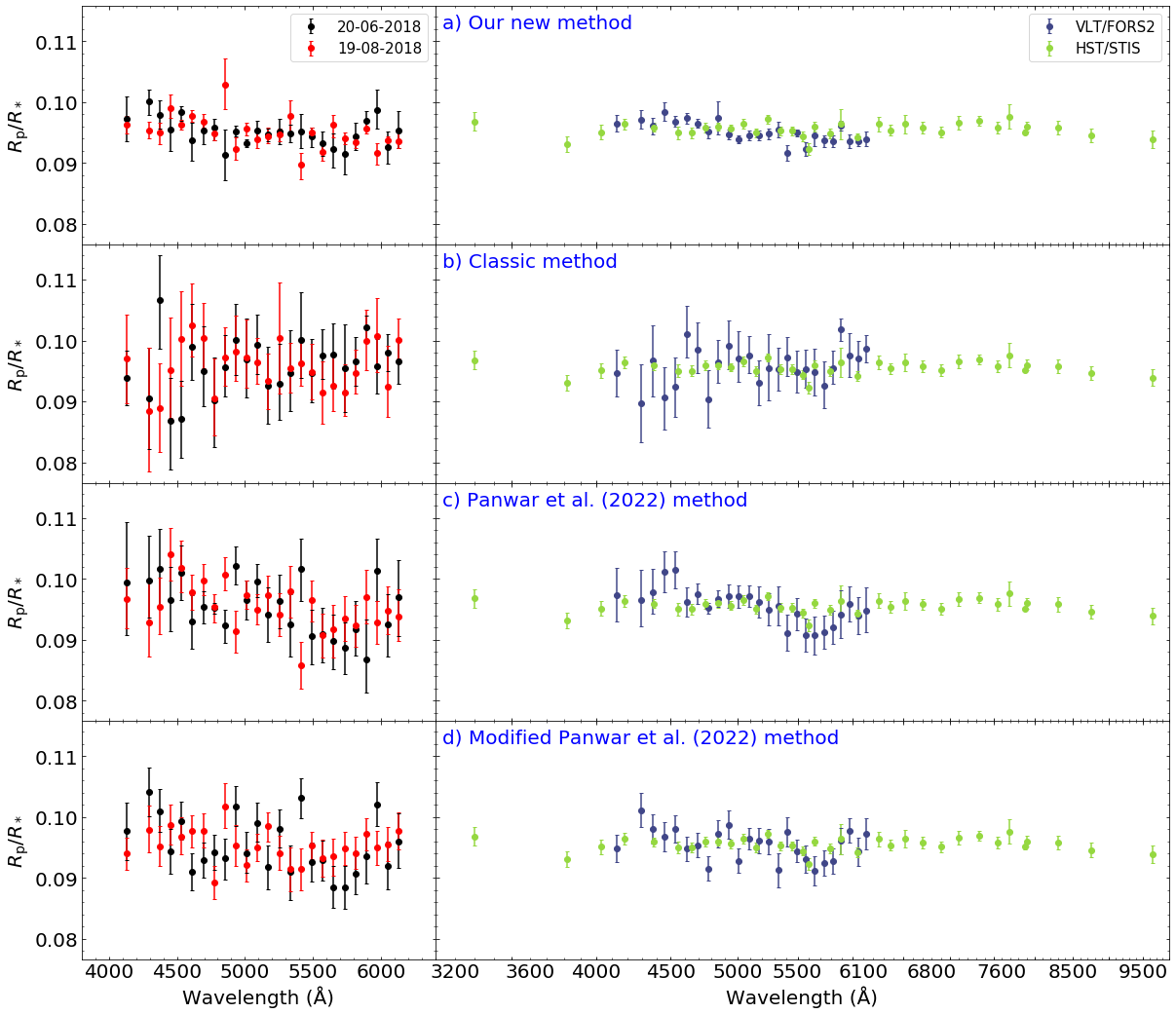}
\caption{\edit{Transmission spectra of WASP-74b from each considered approach, with the individual spectra of each data set on the left and the combined VLT spectra on the right. A constant offset is applied to the June data (black) to bring them to the same level as the August data (red). The combined VLT/FORS2 spectrum (blue) is compared with the result from HST/STIS observations (light green) for each case. From top to bottom, we show results from a) our new method, b) the classic method, c) the \citet{2022MNRAS.510.3236P} method, assuming time and common-mode as GP regressors, and d) the modified \citet{2022MNRAS.510.3236P} method that includes an exponential function of airmass.}}
\label{fig:transmission_spectrum}
\end{figure*}

\begin{figure*}
\centering
\includegraphics[width=\textwidth,height=\textheight,keepaspectratio]{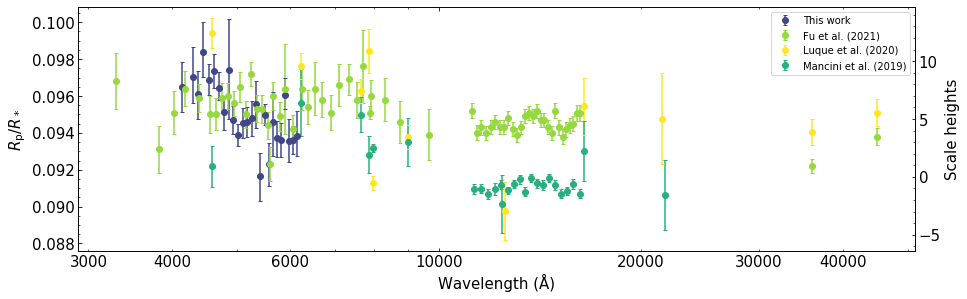}
\caption{\edit{WASP-74b transmission spectrum using data from four studies. The light green circles are from \citet{2021AJ....162..271F}, the yellow circles are from \citet{2020A&A...642A..50L}, and the turquoise circles are from \citet{2019MNRAS.485.5168M}. The blue circles represent the results from this work.}}
\label{fig:transmission_spectrum_2}
\end{figure*}

\section{Results} \label{sec:results}

\subsection{The blue-optical trends of the planetary radius}
\label{sec:transmission_spectrum}

Results from the individual analyses of the June and August data sets, as well as the combined results, are presented in Fig.~\ref{fig:transmission_spectrum}. \edit{A constant offset between the two data sets was determined in all cases}. For our new method detailed in Section~\ref{sec:spectroscopic_analysis_D}, this offset was found to be $\Delta R_\mathrm{p}/R_*$ = 0.0033 $\pm$ 0.0037. We associate this offset with differing observing conditions between the two nights, with June being more affected by turbulence and water vapour and August more affected by clouds (see also Fig.~\ref{fig:optical_state_parameters}). We applied the offset for each respective analysis to the June observation and then calculated the weighted mean $R_\mathrm{p}/R_*$ between the two observations to create the combined transmission spectrum. The produced transmission spectrum shows no evidence of a significant increase in the transit depth and duration at blue wavelengths that could be attributed to atmospheric escape (see Section~\ref{sec:intro}).

Taking a closer look at the transmission spectrum from our new approach, we do not detect any obvious, constant features in the individual transmission spectra that would indicate absorption from atomic or molecular species. In addition, the blue end of the spectrum shows an increasing scattering slope (see also Figs.~\ref{fig:transmission_spectrum_2} and~\ref{fig:platon}) that is comparable to the result found from ground-based photometry \citep{2020A&A...642A..50L}. Regardless of the slope, the average planet-to-star radii ratios are in good agreement with the HST transmission spectrum of \citet{2021AJ....162..271F}. \edit{The \citet{2021AJ....162..271F} spectrum is in partial agreement with retrieval run 2 from \citet{2020A&A...642A..50L}, which ignores the two I-band measurements at $\sim$8000\,\AA. The retrieval of \citet{2020A&A...642A..50L} also includes the WFC3 data, as computed by \citet{2019MNRAS.485.5168M}, but with an offset applied that brings the data closer to \citet{2021AJ....162..271F}. The key difference in the space-based data, apart from the simple Rayleigh scattering slope, is that the \textit{Spitzer} data points are a bit lower with respect to the results from \citet{2020A&A...642A..50L} (see Fig.~\ref{fig:transmission_spectrum_2}). One difference between our analysis and the analysis of \citet{2021AJ....162..271F} is that we assume quadratic limb darkening and allow for a free limb darkening coefficient during the fit, whereas \citet{2021AJ....162..271F} keep all coefficients fixed to theoretical values from the four-parameter law. This change} was important here to mitigate potential effects from the Earth's atmosphere. Fig.~\ref{fig:transmission_spectrum_2} illustrates how this combined transmission spectrum compares against other results from the literature.

\subsection{Atmospheric retrieval using PLATON}
\label{sec:platon}

We utilised the PLATON software \citep{2019PASP..131c4501Z,2020ApJ...899...27Z} to retrieve the atmospheric properties of WASP-74b. PLATON is a versatile code\footnote{https://platon.readthedocs.io/en/latest/} that allows for quick investigations of exoplanetary atmospheres using a wide range of temperatures, metallicities, carbon-to-oxygen (C/O) ratios, cloud-top layers and scattering slopes. PLATON has proven to be a useful tool in constraining the atmospheric properties of several hot exoplanets \citep[e.g.][]{2021MNRAS.500.5420C,2021AJ....161...51S,2021AJ....162...34K,2021A&A...656A.114J,2022MNRAS.510.4857A}.

\begin{figure}
\centering
\includegraphics[width=0.5\textwidth,height=0.4\textheight]{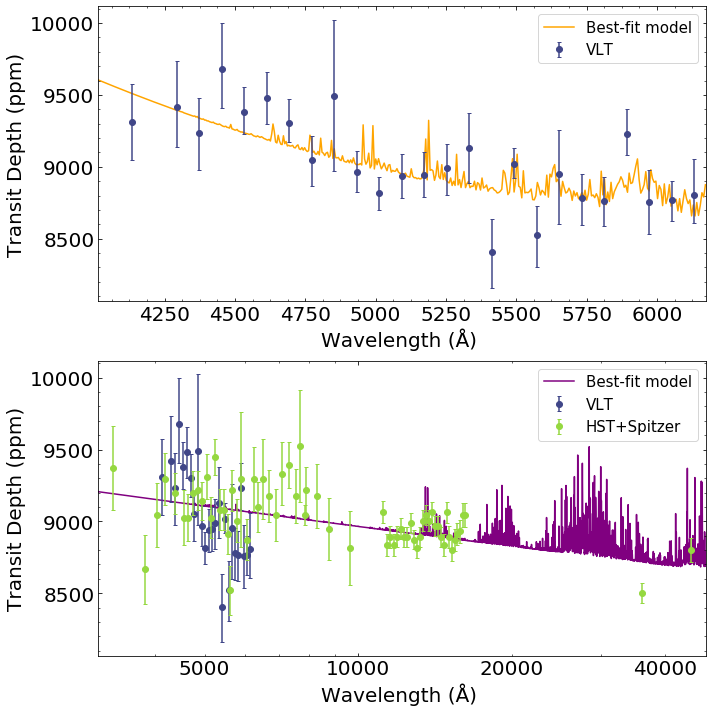}
\caption{\edit{Retrieved transmission spectra of WASP-74b from PLATON for the VLT analysis (top) and the combined VLT+HST+\textit{Spitzer} analysis (bottom).}}
\label{fig:platon}
\end{figure}

We performed two atmospheric retrieval analyses, one that considered only the VLT data and another that also included the published WASP-74b spectroscopic data from \citet{2021AJ....162..271F}. We considered equilibrium chemistry, rainout condensation and an isothermal atmosphere for the planet. The free parameters involved were the planetary radius and temperature, the C/O ratio, the metallicity, two scattering parameters (scattering factor and slope), and one cloud parameter (cloud-top pressure). In addition, we scaled our measurements by a multiplicative factor $\beta$. To get the optimised values for the atmospheric parameters we followed a multimodal nested sampling procedure \edit{\citep[\texttt{dynesty},][]{2020MNRAS.493.3132S}} and used 1000 live points. The transmission spectra from the retrieval processes are shown in Fig.~\ref{fig:platon}, and their posterior distributions can be seen in Fig.~\ref{fig:platon_posterior}, for which details are given in Table~\ref{tab:platon}.

\begin{table}
\centering
\caption{PLATON atmospheric retrieval results for WASP-74b.}
\label{tab:platon}
\begin{tabular}{lccc}
\hline
\hline
Parameter & Prior & VLT & VLT+HST+\textit{Spitzer}\\
\hline
$R_{\rm p}$ (R$_{\rm Jup}$) & $\mathcal{U}$(0.5R$_{\rm p}$, 1.5R$_{\rm p}$) & 1.33 $\pm$ 0.03 & 1.34 $\pm$ 0.03\\[2pt]
$T_{\rm p}$ (K) & $\mathcal{U}$(950, 1900) & $1602^{+193}_{-312}$ & $1310^{+260}_{-228}$\\[2pt]
$\beta$ & $\mathcal{U}$(0.1, 10) & $1.23^{+0.21}_{-0.17}$ & 1.36 $\pm$ 0.10\\[2pt]
log$f_{\rm scatter}$ & $\mathcal{U}$(-4, 10) & $1.1^{+2.3}_{-2.1}$ & $7.4^{+1.2}_{-1.8}$\\[2pt]
log$(Z/Z_{\rm \odot})$ & $\mathcal{U}$(-1, 3) & $0.30^{+0.95}_{-0.83}$ & $1.18^{+1.06}_{-1.37}$\\[2pt]
C/O & $\mathcal{U}$(0.05, 2) & $0.94^{+0.68}_{-0.56}$ & $0.97^{+0.69}_{-0.58}$\\[2pt]
log $P_{\rm cloudtop}$ & $\mathcal{U}$(-3.99, 8) & $4.0^{+2.6}_{-3.2}$ & $3.7^{+2.7}_{-2.8}$\\[2pt]
scatter slope & $\mathcal{U}$(0, 20) & $15.9^{+2.7}_{-4.4}$ & $3.2^{+3.5}_{-1.1}$\\[2pt]
\hline
\end{tabular}
\end{table}

The VLT-specific analysis revealed a scattering slope of $15.9^{+2.7}_{-4.4}$ and a log scattering factor of $1.1^{+2.3}_{-2.1}$. The retrieval result is slightly skewed towards the upper slope limits but is in agreement with a simple scattering slope fit, which gives a scattering slope value of 13.9 $\pm$ 3.4. The upward scattering slope towards shorter wavelengths computed here is comparable to the one estimated by \citet{2020A&A...642A..50L}.

When we incorporate the HST and \textit{Spitzer} data points, we see that our results are closer to the ones from \citet{2021AJ....162..271F}. We find slightly higher values for the scattering slope ($3.2^{+3.5}_{-1.1}$) and the log scattering factor ($7.4^{+1.2}_{-1.8}$) compared to \citet{2021AJ....162..271F}, but both are within 1$\sigma$. This is to be expected given the larger errors obtained in the VLT data sets and the rather restricted bandwidth of these observations. HST STIS/WFC3 and \textit{Spitzer} data explore a much wider spectral range (ultraviolet to infrared) and, hence, they dominate the retrievals towards the atmospheric parameters reported in \citet{2021AJ....162..271F}. Despite that, we observe a degree of consistency between ground and space-based observations with the C/O ratio and the cloud-top pressure \edit{(i.e. the atmospheric layer below which most clouds are expected to form)} settling at similar values. Metallicity, on the other hand, is found to be higher than expected. This result comes naturally as we use broader priors for log$(Z/Z_{\rm \odot})$ compared to \citet{2021AJ....162..271F}. Furthermore, a retrieval assuming an offset for the VLT data set leads to similar results as the offset is small. 

\edit{An additional retrieval analysis assuming a flat (cloudy) model was also conducted for the combined VLT, HST and \textit{Spitzer} data set. In this model, the cloud-top pressure was fixed to 0.001\,Pa and the scattering parameters were fixed to Rayleigh scattering. To compare the hazy model with the flat model we used the log Bayesian evidence values (log$_e\mathcal Z$) computed by \texttt{dynesty}. We found that the log evidence is higher for the hazy model and that the difference $\Delta$log$_e\mathcal Z$ between the two models is greater than 5 ($\Delta$log$_e\mathcal Z$ = 10.65). This result indicates that the hazy model is strongly favoured over the flat model \citep{kass1995bayes}.}

\subsection{Atmospheric retrievals using AURA}
\label{sec:aura}

\begin{figure*}
\centering
\includegraphics[width=\textwidth]{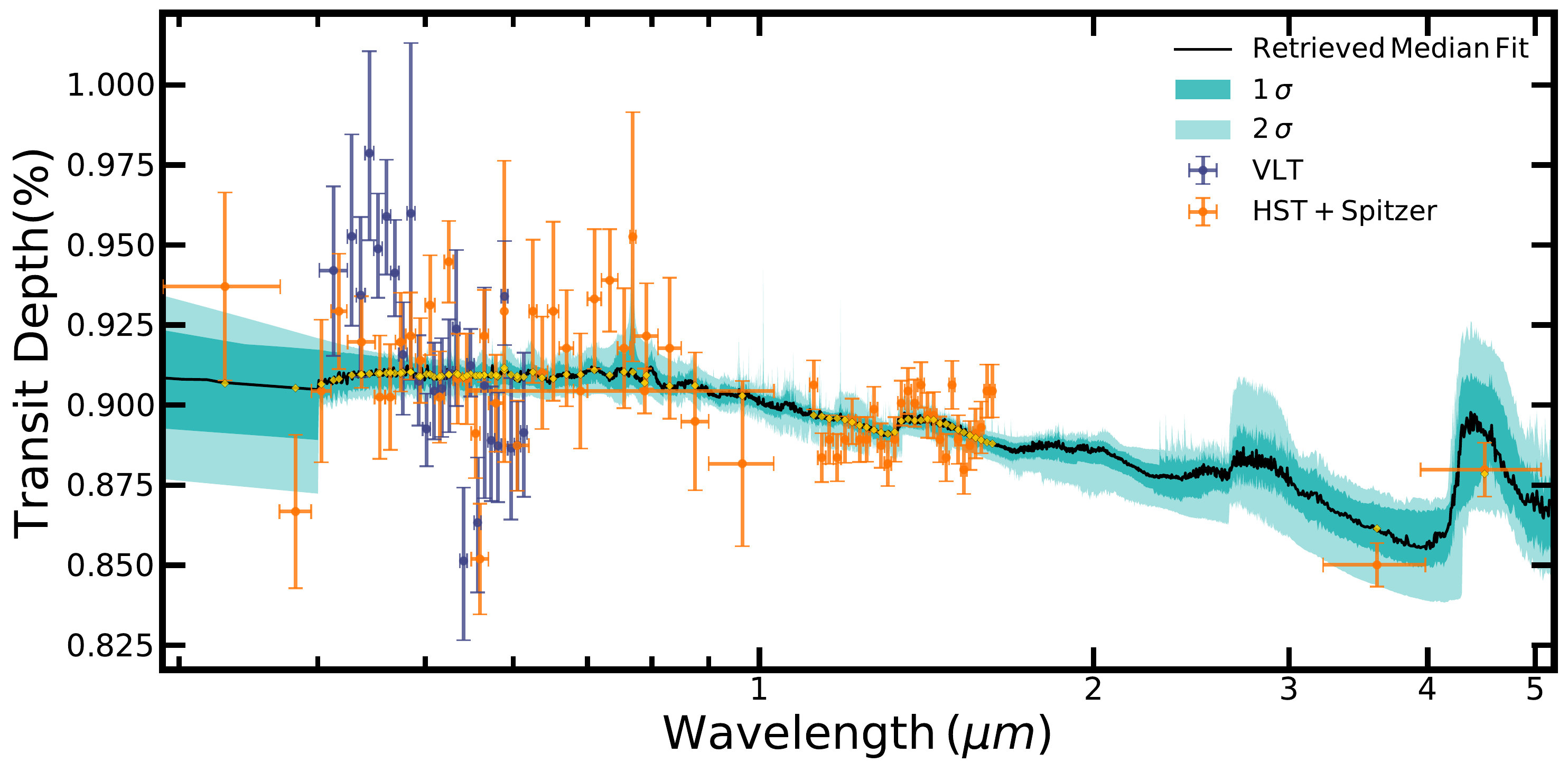}
\caption{Retrieved spectral fit obtained by AURA. The median retrieved spectrum is shown as a black line, while darker and lighter shaded regions denote the corresponding 1$\sigma$ and 2$\sigma$ regions. \edit{Yellow diamonds denote the median retrieved spectrum binned down to match the observations.}}
\label{fig:aura}
\end{figure*}

We also carry out retrievals using the AURA retrieval framework \citep{2018MNRAS.480.5314P}. AURA couples a forward model generator with a robust Bayesian parameter estimator based on the MultiNest nested sampling algorithm \citep{2009MNRAS.398.1601F, 2014A&A...564A.125B}. Forward model transmission spectra are generated by modelling the planet's terminator as a hydrogen-dominated 1D plane-parallel atmosphere in hydrostatic equilibrium. AURA then carries out a radiative transfer calculation, considering opacity contributions from H$_2$-H$_2$ and H$_2$-He collision-induced absorption, as well as several gaseous chemical species. The mixing ratios of all chemical species are free parameters in our model. Clouds are modelled as a grey opacity present at all altitudes below the cloud deck pressure, which is a free parameter. Hazes are modelled as a modification to Rayleigh scattering above the cloud deck, with a cross-section $\sigma = a \sigma_0 (\lambda / \lambda_0)^\gamma$, where $\sigma_0 = 5.31 \times 10^{-31}$~m$^2$ and $\lambda_0 = 350$~nm, while $a$ and $\gamma$ are free parameters. Our model allows for a patchy coverage of the terminator atmosphere by clouds/hazes, given by $\phi$, the coverage fraction.

For the specific case of WASP-74b, we consider models including opacity contributions from H$_2$O \citep{2010JQSRT.111.2139R}, CH$_4$ \citep{2014MNRAS.440.1649Y}, NH$_3$ \citep{2011MNRAS.413.1828Y}, HCN \citep{2014MNRAS.437.1828B}, CO \citep{2010JQSRT.111.2139R} and CO$_2$ \citep{2010JQSRT.111.2139R}, as well as Na and K \citep{2019ApJ...887L..20W}. Given the planet's high temperature, we also include TiO \citep{2019MNRAS.488.2836M} and VO \citep{2016MNRAS.463..771M}, as well as the metal hydrides FeH, CrH and TiH \citep{2001JChPh.115.1312B, 2003ApJ...594..651D, 2005ApJ...624..988B, 2016JMoSp.327...73T}. We additionally carry out retrievals that include AlO \citep{2015MNRAS.449.3613P}, discussing the effect its inclusion has below.

We find our retrieval on the combined WASP-74b observations produces a spectral fit with muted chemical absorption features, while not producing any precise constraints for the composition of the atmosphere\edit{, as shown in Fig.~\ref{fig:aura_posterior}}. Specifically, our retrieval produces a posterior distribution for the mixing ratio of H$_2$O that is peaked but largely unconstrained, with a 2$\sigma$ upper limit of $10^{-2.5}$\edit{, corresponding to a $\sim$10$\times$ solar elemental oxygen abundance}. The same is seen for the mixing ratio posterior distributions of K and CO, with corresponding 2$\sigma$ upper limits of $10^{-2.9}$ and $10^{-1.8}$, respectively. \edit{ The upper limits are equivalent to a $\sim$10,000$\times$ solar elemental enhancement for K and $\sim$50$\times$ solar for C.}

Our retrievals also find a strong spectral contribution from hazes. This gives rise to the scattering slope seen in Fig.~\ref{fig:aura}, which decreases with wavelength to reach the two \textit{Spitzer} photometric points that lie lower relative to the other observations. Our retrieval constrains the Rayleigh enhancement factor, $\mathrm{log}(a)$, to $7.2^{+1.6}_{-2.3}$ and the scattering slope $\gamma$ to $-7.2^{+2.5}_{-1.9}$. It also finds that hazes partially cover the terminator atmosphere, with a coverage fraction of $0.6^{+0.2}_{-0.2}$.

For our retrieval that includes AlO, we once again find a strong spectral contribution from Rayleigh-like hazes, retrieving haze parameter constraints that are largely consistent with those obtained with our other retrieval. Specifically, $\mathrm{log}(a)$ is constrained to $5.5^{+1.8}_{-1.3}$, while $\gamma$ is constrained to $-13.7^{+6.1}_{-4.2}$ and the coverage fraction to $0.6^{+0.2}_{-0.2}$. Additionally, the mixing ratio of AlO is constrained to an unrealistically high value of $\sim 10^{-3.5}$. While metal oxides such as AlO are expected to be present in the atmosphere of a high-temperature planet like WASP-74b, the retrieved mixing ratio is \edit{$\sim$ 5} dex higher than equilibrium expectations\edit{ for solar elemental abundances. This is due to Al additionally being present in other more abundant refractory species under chemical equilibrium \citep{2018AA...614A...1W}, thereby requiring a significant elemental enhancement to give rise to high AlO mixing ratios}. The retrieved spectral fit displays numerous small AlO absorption features in the optical, which are partially masked by the haze spectral contributions. Given the unphysically high retrieved mixing ratio and the nature of its resulting spectral contributions, it is likely that the retrieved AlO constraints are driven by small noise features.

\section{Discussion}

\subsection{The benefits and drawbacks of our new method}

The novel approach we presented in the spectroscopic analysis is based on the raw flux of WASP-74 and therefore eliminates the additional scatter that comes with taking the comparison star into account. In that respect, this method is similar to \citet{2022MNRAS.510.3236P}, where the fit is also performed on the raw light curves of the target. This is also the reason why both methods show a significant reduction in the light curve scatter, reaching almost 80\% of the photon noise limit.

Another key element of our new approach is the use of an exponential of airmass that models extinction effects. While parametric functions of airmass have been applied to target-to-reference star fluxes before \citep[e.g.][]{2015A&A...576L..11S}, here we isolate the target and apply such a function to the spectroscopic flux of the target only. We found that the inclusion of this parametric function in the spectroscopic light curve models improves the fit considerably in both the \citet{2022MNRAS.510.3236P} method and our own novel approach and offers enhanced precision. At the same time, the produced transmission spectra have a similar characteristic shape, showcasing the reliability of our new method.

Not only that, but our new method provides remarkably low uncertainties that are comparable to the HST space-based results. This outcome is more extraordinary considering that, in this study, we use a more conservative approach based on GPs. We attribute this improvement to the smoother shape of the common-mode corrected light curves. In the \citet{2022MNRAS.510.3236P} recipe, the GP tries to account for systematic effects in the raw spectroscopic light curves using both time and common-mode as regressors. This parameterisation is applied to raw fluxes and may have more difficulties in evaluating discontinuities caused by unaccounted for systematics and other effects such as cloud crossings. In our new method, the noise contribution from aerosols is greatly corrected for linearly during common-mode correction owing to the relatively homogeneous distribution of the clouds in the June and August observations.

Despite its success in generating a credible and precise transmission spectrum, our new method depends on the curvature of the light curves due to the inclusion of the airmass exponential. This means that our method is limited to ground-based data, where the Earth's atmosphere plays a fundamental role in the observed shape of the light curves. Furthermore, the use of this model is less effective for light curves that demonstrate linear trends.

Another limiting factor is the common-mode correction itself, which can potentially cause a systematic ``domino effect'' from one spectroscopic light curve to another. For example, a cosmic ray affecting certain wavelengths or an increased scatter in parts of the detector, due to detector cosmetics, can influence other light curves by introducing systematics from one light curve into another. This is because the common-mode assumption may describe white noise for the most part but can be affected by certain systematic trends in specific wavelength regions. If these systematic effects are dominant, then they can introduce additional interference in other parts of the spectrum. Our outlier removal algorithm may correct for cosmic rays but detector artifacts can be trickier and may be present in multiple data points or the entire observations.

Traditionally, the use of parametric functions meant that a comparison star was the only way to correct for atmospheric effects due to the inflexibility of these functions to reliably fit light curves that exhibit unknown systematics. This is also evident from the Levenberg-Marquardt fits that show inconsistencies in some of the light curves and is one of the reasons why these fits were only used to obtain initial guesses. The emergence of GPs, however, has challenged this approach, because GPs can take into account unaccounted for systematics. Both the \citet{2022MNRAS.510.3236P} approach and this new method rely on this notion to construct a convincing transmission spectrum.

We note that the improvement in the transit depth presented in this study was found for blue-optical data for a specific target and may vary for other wavelengths and other targets. \edit{We also stress that out-of-transit data is available for our study, which facilitates the determination of the exponential coefficients. Data sets with a limited number of out-of-transit data points may not be well-suited for this new approach.} In addition, while the comparison star is not considered in the spectroscopic light curves, it is still being employed in the white light curve analysis to obtain the transit model used in the common-mode correction. This also differentiates our approach from \citet{2022MNRAS.510.3236P} who removed the need for a comparison star entirely. Nonetheless, our new method provides a valuable and useful alternative that can immensely aid the characterisation of exoplanetary atmospheres providing increased precision. 

\subsection{WASP-74b in context}
\label{sec:context}

WASP-74b has an equilibrium temperature that places it in the transitional region between hot Jupiters and ultra-hot Jupiters ($\sim$1500-2000\,K). This is a temperature range that has not yet been studied in detail and when the surface gravity of WASP-74b is taken into account, the sample of investigated hot Jupiter atmospheres becomes quite small. CoRoT-1b ($g_{\rm p}$=10.65\,m\,s$^{-2}$, $T_{\rm eq}$=1915\,K, \citealp{2011MNRAS.417.2166S}), and WASP-79b ($g_{\rm p}$=8.39\,m\,s$^{-2}$, $T_{\rm eq}$=1716\,K, \citealp{2017MNRAS.464..810B}) are two exoplanets with similar bulk characteristics that are part of this sample and have been observed in low resolution. CoRoT-1b has mostly been observed in the infrared, with data revealing a featureless spectrum and indicating an atmosphere obscured by clouds \citep{2014ApJ...783....5S,2014ApJ...785..148R,2022AJ....164...19G}. WASP-79b, on the other hand, is more intriguing as it seems to have an inverted slope towards bluer wavelengths. Such a characteristic may be indicative of stellar contamination from unocculted faculae \citep{2021AJ....162..138R}. In addition, H$_2$O, and possibly H$^-$ and FeH, were also detected \citep{2020AJ....159....5S,2020AJ....160..109S,2021AJ....162..138R}. These findings make WASP-74b the only planet of the group with signs of strong scattering within its atmosphere.

An enhanced scattering slope towards bluer wavelengths is not unusual and is progressively being observed in more exoplanetary atmospheres \citep[e.g.][]{2013MNRAS.432.2917P,2020AJ....160...51A,2020MNRAS.497.5182A,2021MNRAS.500.5420C,2022MNRAS.510.4857A}. While such slopes can be associated with stellar heterogeneity \citep[e.g.][]{2014ApJ...791...55M}, there is currently no indication that the star is active \citep{2015AJ....150...18H,2021AJ....162..271F}. It is, therefore, more likely that the result is influenced by physical processes within the planetary atmosphere and/or contamination from unknown noise sources. If we consider that the effect is intrinsic to the planetary atmosphere, then photochemical processes or mineral condensation could play a role. For example, the formation of hydrocarbon hazes due to photochemical reactions could produce slopes in the transmission spectra \citep{2019ApJ...877..109K,2020ApJ...895L..47O}, but the equilibrium temperature of WASP-74b is somewhat higher than the reported maximum limit of 1500\,K. Furthermore, sulphide species, such as manganese sulphide, could condense at high altitudes to form clouds. Manganese sulphide can produce very steep slopes but tends to form condensates at lower temperatures \citep{2017MNRAS.471.4355P}. Other candidates include silicate species, such as enstatite, and alumina. These species can form mineral clouds at temperatures closer to the retrieved temperature of WASP-74b, but the produced slope tends to be less steep. \citet{2020NatAs...4..951G} found that silicate aerosols are likely dominant at temperatures above $\sim$950K and that cloud formation due to iron and metal sulphides is largely inhibited by low nucleation energies.

\section{Summary}
\label{sec:summary}

Previous works to understand the atmosphere of WASP-74b have so far been inconclusive leading to very conflicting outcomes, with \citet{2019MNRAS.485.5168M} tentatively hinting at molecular absorbers in the atmosphere, \citet{2020A&A...642A..50L} showing potential evidence for increased, super-Rayleigh scattering, and \citet{2021AJ....162..271F} indicating Rayleigh scattering that extends well into the infrared. \citet{2021AJ....162..271F} also performed an eclipse retrieval analysis finding an overall featureless spectrum in the infrared and possible methane absorption based mostly on a single \textit{Spitzer} 3.6\,$\mu$m data point.

In this study, we presented transmission spectroscopy results from observations obtained using the ground-based VLT FORS2 instrument. The data were collected using the dispersive element GRIS600B and we explored the blue-optical wavelengths for three nights. The first night was subsequently rejected due to cloudy weather affecting most of the observation. 

We analysed the two remaining data sets in combination and considered a series of different methodologies in the spectroscopic analysis in our effort to reduce scatter and achieve higher precision. We ultimately developed a new method that bypasses the necessity to include reference stars in the spectrophotometric light curve fits. The new method is similar to the novel approach presented by \citet{2022MNRAS.510.3236P} and has two main characteristics: 1) common-mode correction is applied on the raw spectroscopic light curves of the target, and 2) the effects of extinction are modelled out during the fits with the aid of an exponential function of airmass. We found that this technique improved the fit considerably by minimising uncertainties and providing reliable results.

Our analysis did not reveal a substantially higher planetary radius at the blue end of the optical spectrum and so found no signs of strong absorption at those wavelengths. This suggests that an evaporating atmosphere is unlikely and that the observations in the U-band by \citet{2019MNRAS.485.5168M} are in all likelihood affected by unstable weather conditions during the time of the observations, as acknowledged by those authors.

A retrieval analysis with PLATON to the VLT result, based on equilibrium chemistry, revealed an enhanced scattering slope that is very similar to the one reported by \citet{2020A&A...642A..50L}. However, the steepness of the slope was reduced when additional HST and \textit{Spitzer} data were incorporated in the retrieval. This result is not that surprising considering that the space-based spectrum is more extended and advocates the presence of clouds \citep{2021AJ....162..271F}. Even so, the scattering parameters in the combined case were found to be slightly increased compared to the findings from \citet{2021AJ....162..271F} but the difference was determined to be smaller than 1$\sigma$.

We then conducted retrievals on the combined VLT, HST and \textit{Spitzer} data set using the AURA retrieval framework. This framework provides a somewhat different view of the planets' atmospheric structure in the terminator region and allows for free chemistry. This distinction enables AURA to explore more atmospheric properties, including the abundances of alkali metals, water, metal oxides and metal hydrides. We found that the broad spectral retrievals with AURA also favour an enhanced scattering slope suggestive of haze in the atmosphere of WASP-74b. Interestingly, when we include AlO in the retrievals, we obtain an AlO mixing ratio that is several orders of magnitude higher than expected. The temperature of this planet could favour the presence of this mineral, but the \edit{unrealistically high abundance retrieved} likely suggests minor noise contributions.

Despite the substantial corruption from systematic effects, we managed to obtain a relatively precise transmission spectrum. Additional spectroscopic observations in the blue, as well as complimentary observations in the red and near-infrared will offer a better understanding of the hot Jupiter's atmosphere and will help clear up the picture for this transiting hot Jupiter. The newly commissioned \edit{JWST} will be able to spectroscopically observe the redder wavelengths beyond 0.6\,$\mu$m and will, therefore, immensely help efforts to decipher the properties of WASP-74b.

\section*{Acknowledgements}
This work is based on observations collected at the European Organization for Astronomical Research in the Southern Hemisphere under the European Southern Observatory programme 0101.C-0716. PS is supported by a UK Science and Technology Facilities Council (STFC) studentship. This project has received funding from the European Research Council (ERC) under the European Union’s Horizon 2020 research and innovation programme (project {\sc Four Aces}; grant agreement No 724427). It has also been carried out in the frame of the National Centre for Competence in Research PlanetS supported by the Swiss National Science Foundation (SNSF). DE acknowledges financial support from the Swiss National Science Foundation for project 200021\_200726. L.M. acknowledges support from the ``Fondi di Ricerca Scientifica d'Ateneo 2021'' of the University of Rome ``Tor Vergata''.

\section*{Data Availability}

The VLT FORS2 data are publicly available on the ESO archive under programme 0101.C-0716 (P.I.\ Southworth).

\bibliographystyle{mnras}
\bibliography{references}

\appendix

\section{Atmospheric retrieval distributions}

Here we present the corner plots for the posterior distributions of WASP-74b from the atmospheric retrievals using PLATON and AURA.

\begin{figure*}
\centering
\includegraphics[width=\textwidth,height=\textheight,keepaspectratio]{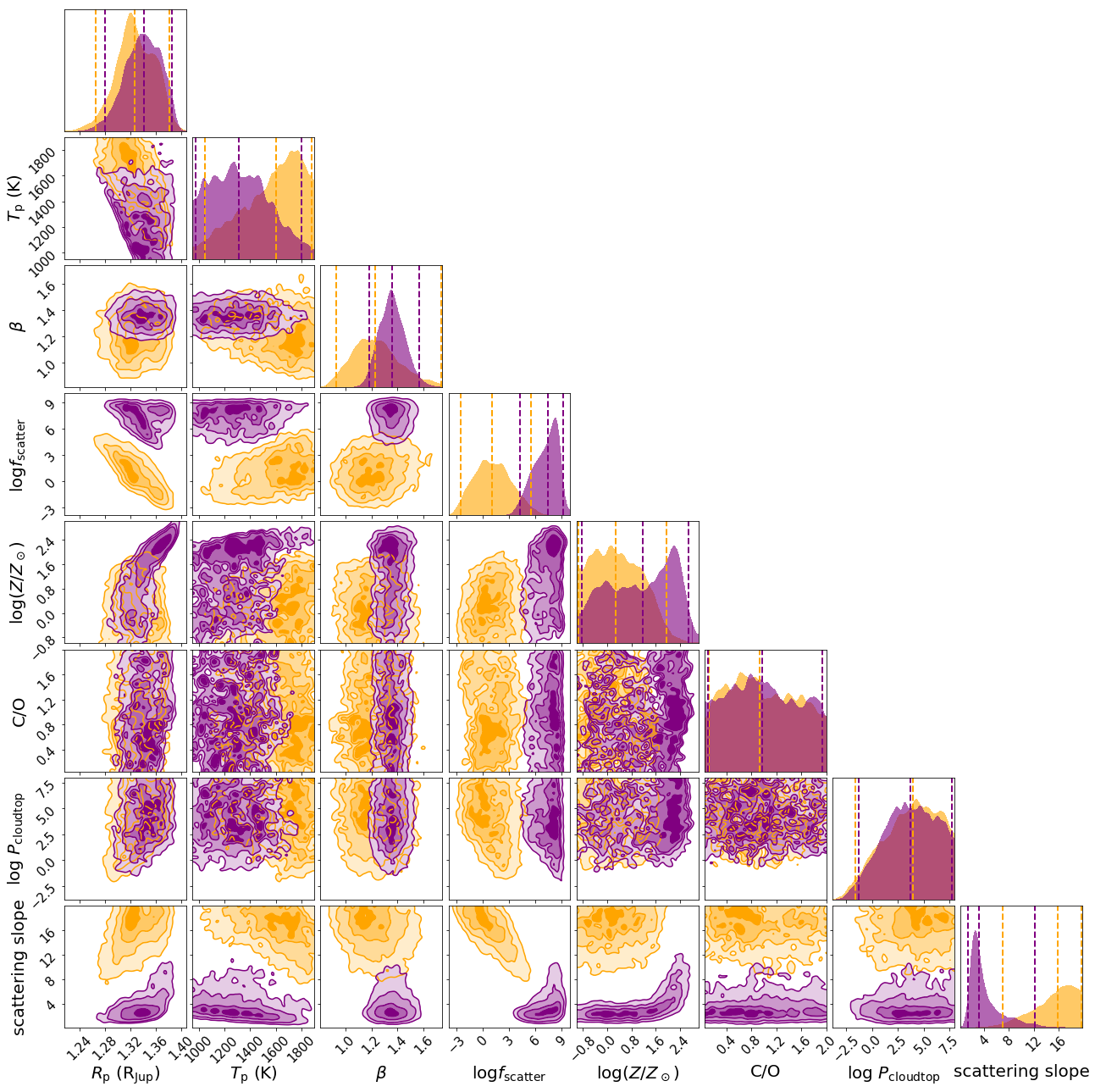}
\caption{\edit{The PLATON atmospheric retrieval distributions of WASP-74b using the combined VLT data set (orange) and the combined VLT+HST+\textit{Spitzer} data set (purple).}}
\label{fig:platon_posterior}
\end{figure*}

\begin{figure*}
\centering
\includegraphics[width=\textwidth,height=\textheight,keepaspectratio]{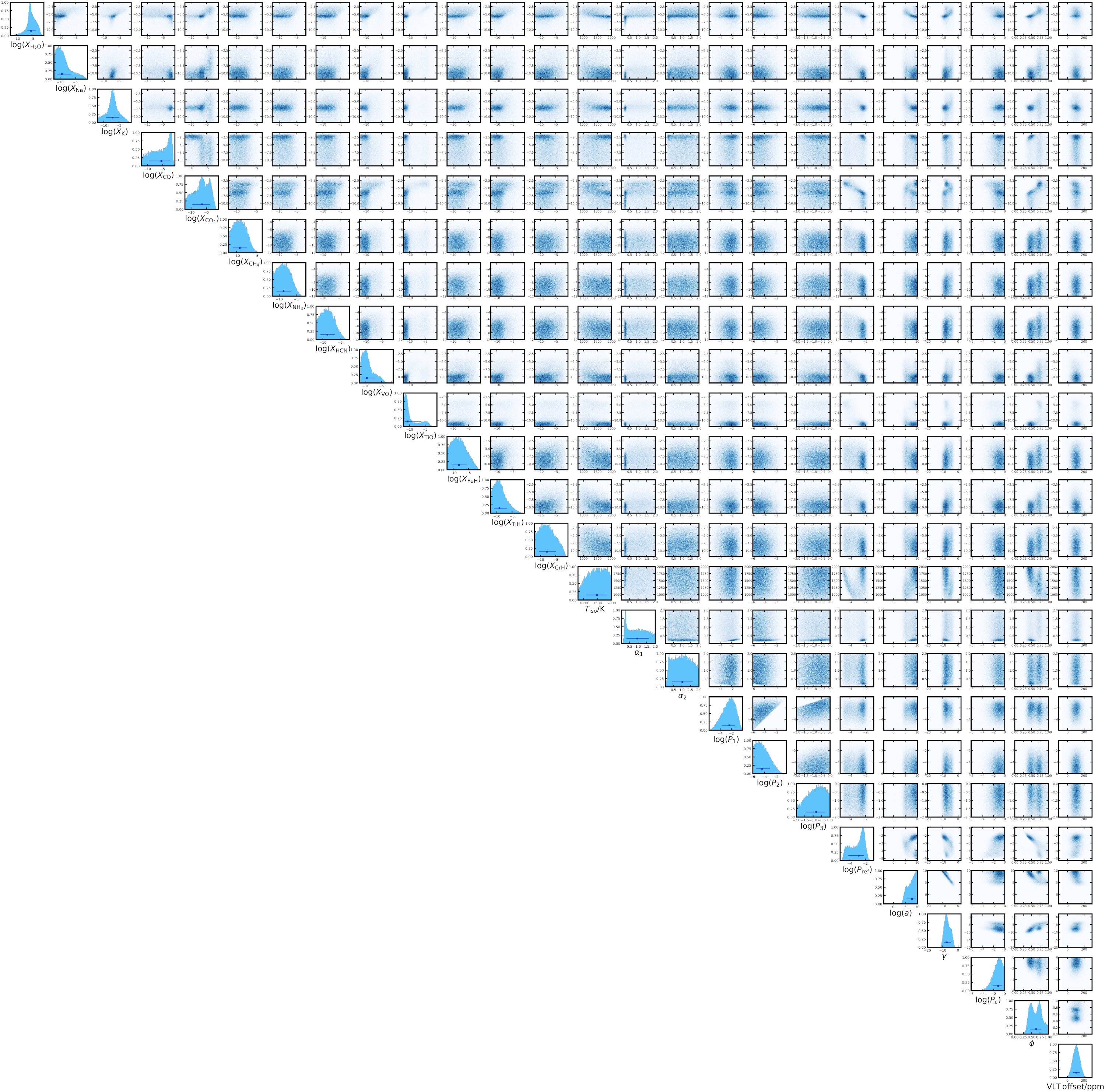}
\caption{The retrieved posterior distribution obtained with AURA using the combined VLT+HST+\textit{Spitzer} dataset.}
\label{fig:aura_posterior}
\end{figure*}

\section{Optical State Parameters}

We also include a visual representation of the behaviour of some optical state parameters throughout the transit observations in June and August 2018.

\begin{figure*}
\centering
\includegraphics[width=\textwidth,height=0.9\textheight,keepaspectratio]{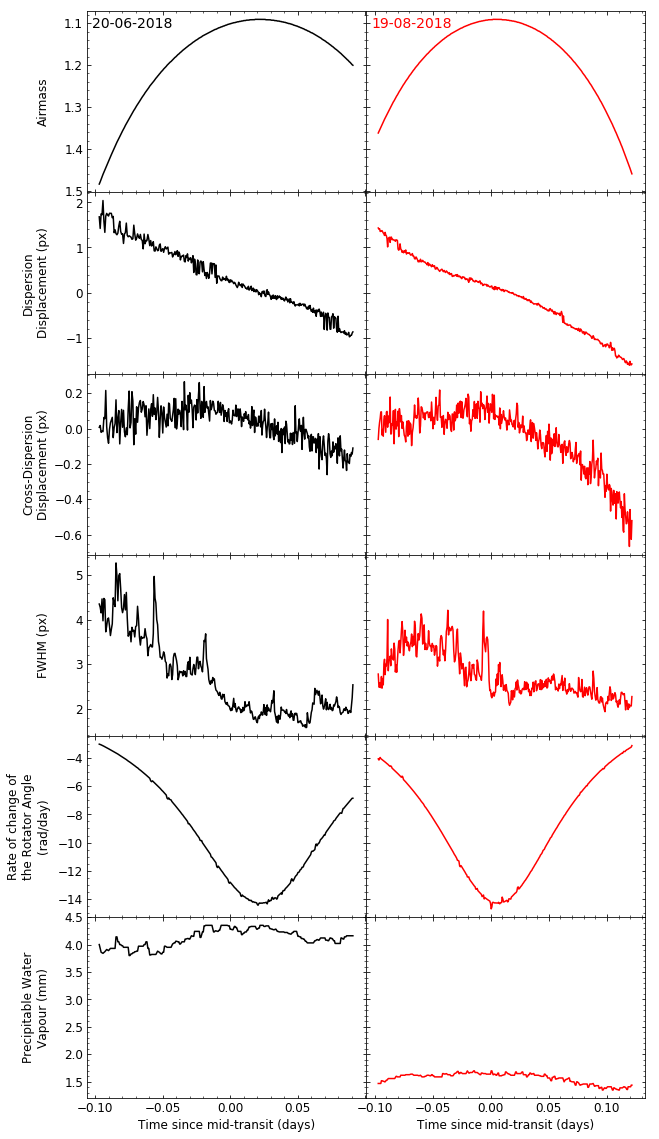}
\caption{Auxiliary variables with respect to time for the two data sets analysed. The displayed parameters, from top to bottom, are airmass, shifts in the dispersion and cross-dispersion axes, FWHM, the rate of change of the rotator angle, and the integrated water vapour.}
\label{fig:optical_state_parameters}
\end{figure*}

\bsp	
\label{lastpage}
\end{document}